\documentclass{aa}
\usepackage{graphicx}
\usepackage{amssymb,amsmath}
\usepackage{latexsym}
\usepackage{mathtools}
\usepackage{epsfig}
\usepackage{natbib}
\usepackage{txfonts}
\begin{document}
   \title{The future large obliquity of Jupiter}
%
   \author{Melaine Saillenfest\inst{1}
          \and
          Giacomo Lari\inst{2}
          \and
          Ariane Courtot\inst{1}
          }
   \authorrunning{Saillenfest et al.}
   \institute{IMCCE, Observatoire de Paris, PSL Research University, CNRS, Sorbonne Universit\'e, Universit\'e de Lille, 75014 Paris, France
              \email{melaine.saillenfest@obspm.fr}
              \and
              Department of Mathematics, University of Pisa, Largo Bruno Pontecorvo 5, 56127 Pisa, Italy
             }
   \date{Received 18 May 2020 / Accepted 3 June 2020}


  \abstract
  {}
  {We aim to determine whether Jupiter's obliquity is bound to remain exceptionally small in the Solar System, or if it could grow in the future and reach values comparable to those of the other giant planets.}
  {The spin axis of Jupiter is subject to the gravitational torques from its regular satellites and from the Sun. These torques evolve over time due to the long-term variations of its orbit and to the migration of its satellites. With numerical simulations, we explore the future evolution of Jupiter's spin axis for different values of its moment of inertia and for different migration rates of its satellites. Analytical formulas show the location and properties of all relevant resonances.}
  {Because of the migration of the Galilean satellites, Jupiter's obliquity is currently increasing, as it adiabatically follows the drift of a secular spin-orbit resonance with the nodal precession mode of Uranus. Using the current estimates of the migration rate of the satellites, the obliquity of Jupiter can reach values ranging from $6^\circ$ to $37^\circ$ after $5$~Gyrs from now, according to the precise value of its polar moment of inertia. A faster migration for the satellites would produce a larger increase in obliquity, as long as the drift remains adiabatic.}
  {Despite its peculiarly small current value, the obliquity of Jupiter is no different from other obliquities in the Solar System: It is equally sensitive to secular spin-orbit resonances and it will probably reach comparable values in the future.}


   \maketitle

\section{Introduction}
   The obliquity of a planet is the angle between its spin axis and the normal to its orbit. A non-zero obliquity results in seasonal climate changes along the planet's orbit, as occurs on Earth. In the protoplanetary disc, giant planets are expected to form with near-zero obliquities, while terrestrial planets should exhibit more random values (see e.g. \citealp{Ward-Hamilton_2004,Rogoszinski-Hamilton_2020a}). Yet, the planets of the Solar System all feature a large variety of obliquities. The case of Mercury is special because the strong tidal dissipation due to the proximity of the Sun now tightly maintains Mercury's obliquity to a near-zero value (see e.g. \citealp{Correia-Laskar_2010}). Excluding Mercury, Jupiter is by far the planet of the Solar System that has the smallest obliquity (see Table~\ref{tab:curob}). This small value seems to put Jupiter in a different category, and it appears unclear why Jupiter should be the only giant planet to indefinitely preserve its primordial obliquity.
   
   Large obliquity changes can be produced by strong impacts. An impact with a planetary-sized body is thought to have created the Moon and affected the spin axis of the Earth, which has remained unchanged ever since \citep{Canup-Asphaug_2001,Li-Batygin_2014b}. Large-scale collisions have also probably participated in increasing the obliquity of Uranus \citep{Boue-Laskar_2010,Morbidelli-etal_2012,Rogoszinski-Hamilton_2020a}.
   
   \begin{table}
      \caption{Current obliquities of the planets of the Solar System.}
      \label{tab:curob}
      \vspace{-0.7cm}
      \begin{equation*}
         \begin{array}{rrcrr}
         \hline
         \hline
                        & \text{obliquity} &&                & \text{obliquity} \\
         \hline
         \text{Mercury} &       0.03^\circ && \text{Jupiter} &   3.12^\circ \\
           \text{Venus} &     177.36^\circ &&  \text{Saturn} &  26.73^\circ \\
           \text{Earth} &      23.45^\circ &&  \text{Uranus} &  97.86^\circ \\
            \text{Mars} &      25.19^\circ && \text{Neptune} &  29.56^\circ \\
         \hline
         \end{array}
      \end{equation*}
      \vspace{-0.5cm}
      \tablefoot{Mercury's obliquity is taken from \cite{Konopliv-etal_2020}. Other values are taken from \cite{Murray-Dermott_1999} who cite the compilation made by \cite{Yoder_1995}.}
   \end{table}  
   
   Apart from collisions, a well-known mechanism that can modify the obliquity of a planet is a so-called ``secular spin-orbit resonance'', that is, a near commensurability between the frequency of precession of the spin axis and the frequency of one (or several) harmonics appearing in the precession of the orbit. This mechanism happens to be extremely common in planetary systems. The overlap of such resonances produces a large chaotic region for the spin axis of the terrestrial planets of the Solar System (see \citealp{Laskar-Robutel_1993}). This chaos probably had a strong influence on the early obliquity of Venus, which was then driven to its current value by the solar tides combined with its thick atmosphere \citep{Correia-Laskar_2001,Correia-etal_2003,Correia-Laskar_2003}. The Moon currently protects the Earth from large chaotic variations in its obliquity \citep{Laskar-etal_1993a,Li-Batygin_2014a}, but due to tidal dissipation within the Earth-Moon system, the Earth will eventually reach the chaotic region in a few gigayears from now (see \citealp{NerondeSurgy-Laskar_1997}). This chaotic zone also strongly affects the obliquity of Mars, which still currently wanders between $0^\circ$ and more than $60^\circ$ \citep{Laskar-etal_2004a,Brasser-Walsh_2011}. As shown by \cite{Millholland-Batygin_2019}, secular spin-orbit resonances can also take place very early in the history of a planet, that is, within the protoplanetary disc itself. More generally, secular spin-orbit resonances are thought to strongly affect the obliquity of exoplanets (see e.g. \citealp{Atobe-etal_2004,Brasser-etal_2014,Deitrick-etal_2018a,Deitrick-etal_2018b,Shan-Li_2018,Millholland-Laughlin_2018,Millholland-Laughlin_2019,Quarles-etal_2019,Saillenfest-etal_2019a,Kreyche-etal_2020}).
   
   For the giant planets of the Solar System, the secular spin-orbit resonances are relatively thin today and well separated from each other. This is why it is so difficult to explain the large obliquity of Uranus by a spin-orbit coupling, now that the precession of Uranus' spin axis is far from any first-order resonances (see e.g. \citealp{Boue-Laskar_2010}, \citealp{Rogoszinski-Hamilton_2020a,Rogoszinski-Hamilton_2020b}). Jupiter and Saturn, on the contrary, are located very close to strong resonances: Jupiter is close to resonance with the nodal precession mode of Uranus \citep{Ward-Canup_2006}, and Saturn is close to resonance with the nodal precession mode of Neptune \citep{Ward-Hamilton_2004,Hamilton-Ward_2004,Boue-etal_2009}. Therefore, the dynamics of Jupiter's spin axis seems to be equally affected by secular spin-orbit resonances as other planets in the Solar System. This was confirmed by \cite{Brasser-Lee_2015} and \cite{Vokrouhlicky-Nesvorny_2015}, who show that models of the late planetary migration have to be finely tuned to avoid overexciting Jupiter's obliquity by spin-orbit coupling, while tilting Saturn to its current orientation. In this regard, the spin-axis dynamics of Jupiter does not appear to be special at all, in contrast to its small obliquity value.
   
   In this article, we aim to investigate the future long-term spin-axis dynamics of Jupiter. In particular, we want to determine whether Jupiter's obliquity is bound to remain exceptionally small in the Solar System, or if it could grow in the future and reach values comparable to those of the other planets.
   
   The precession motion of a planet's spin axis depends on the physical properties of the planet (mass repartition and spin velocity), but also on external torques applied to its equatorial bulge. These torques come from the combined gravitational attraction of the Sun and of satellites (if it has any). Since the orbit of Jupiter is stable over billions of years \citep{Laskar_1990}, the direct torque from the Sun will not noticeably change in the future. However, Jupiter's satellites are known to migrate over time because of tidal dissipation. The future long-term orbital evolution of the Galilean satellites has been recently explored by \cite{Lari-etal_2020}. The solutions that they describe can therefore be used as a guide to study the future spin-axis dynamics of Jupiter. Due to their much smaller masses, the other satellites of Jupiter do not contribute noticeably to its spin-axis dynamics.
   
   In Sect.~\ref{sec:dyn}, we describe our dynamical model and discuss the range of acceptable values for the physical parameters of Jupiter, in particular its polar moment of inertia. In Sect.~\ref{sec:evol}, we present our results about the future spin-axis dynamics of Jupiter: We explore the outcomes given by different values of the poorly known physical parameters of Jupiter and by different migration rates for its satellites. Our conclusions are summarised in Sect.~\ref{sec:ccl}.
   
\section{Secular dynamics of the spin axis}\label{sec:dyn}

   \subsection{Equations of motion}\label{ssec:eqmot}
   The spin-axis dynamics of an oblate planet subject to the lowest-order term of the torque from the Sun is given for instance by \cite{Laskar-Robutel_1993} or \cite{NerondeSurgy-Laskar_1997}. Far from spin-orbit resonances, and due to the weakness of the torque, the long-term evolution of the spin axis is accurately described by the secular Hamiltonian function (i.e. averaged over rotational and orbital motions). This Hamiltonian can be written
   \begin{equation}\label{eq:Hinit}
      \begin{aligned}
         \mathcal{H}(X,-\psi,t) &= -\frac{\alpha}{2}\frac{X^2}{\big(1-e(t)^2\big)^{3/2}} \\
         &- \sqrt{1-X^2}\big(\mathcal{A}(t)\sin\psi + \mathcal{B}(t)\cos\psi\big) \\
         &+ 2X\mathcal{C}(t),
      \end{aligned}
   \end{equation}
   where the conjugate coordinates are $X$ (cosine of obliquity) and $-\psi$ (minus the precession angle). The Hamiltonian in Eq.~\eqref{eq:Hinit} depends explicitly on time $t$ through the orbital eccentricity $e$ and through the functions
   \begin{equation}
      \left\{
      \begin{aligned}
         \mathcal{A}(t) &= \frac{2\big(\dot{q}+p\,\mathcal{C}(t)\big)}{\sqrt{1-p^2-q^2}}\,, \\
         \mathcal{B}(t) &= \frac{2\big(\dot{p}-q\,\mathcal{C}(t)\big)}{\sqrt{1-p^2-q^2}} \,,\\
      \end{aligned}
      \right.
      \quad
      \text{and}
      \quad
      \mathcal{C}(t) = q\dot{p}-p\dot{q}\,.
   \end{equation}
   In these expressions, $q=\eta\cos\Omega$ and $p=\eta\sin\Omega$, where $\eta\equiv\sin(I/2)$, and $I$ and $\Omega$ are the orbital inclination and the longitude of ascending node of the planet, respectively. If the orbit of the planet is fixed in time, its obliquity is constant and its precession angle $\psi$ circulates with constant angular velocity $\alpha X/(1-e^2)^{3/2}$. The quantity $\alpha$ is called the precession constant. It depends on the spin rate of the planet and of its mass distribution, through the formula:
   \begin{equation}\label{eq:alpha}
      \alpha = \frac{3}{2}\frac{\mathcal{G}m_\odot}{\omega a^3}\frac{J_2}{\lambda} \,,
   \end{equation}
   where $\mathcal{G}$ is the gravitational constant, $m_\odot$ is the mass of the Sun, $\omega$ is the spin rate of the planet, $a$ is its semi-major axis, $J_2$ is its second zonal gravity coefficient, and $\lambda$ is its normalised polar moment of inertia. We retrieve the expression given for instance by \cite{NerondeSurgy-Laskar_1997} by noting that
   \begin{equation}\label{eq:J2lb}
      J_2 = \frac{2C-A-B}{2MR_\mathrm{eq}^2}
      \quad\text{and}\quad
      \lambda = \frac{C}{MR_\mathrm{eq}^2}\,,
   \end{equation}
   where $A$, $B$, and $C$ are the equatorial and polar moments of inertia of the planet, $M$ is its mass, and $R_\mathrm{eq}$ is its equatorial radius.
   
   The precession rate of the planet is increased if it possesses massive satellites. If the satellites are far away from the planet, their equilibrium orbital plane (called Laplace plane, see \citealp{Tremaine_2009}) is close to the orbital plane of the planet; therefore, far-away satellites increase the torque exerted by the Sun on the equatorial bulge of the planet. If the satellites are close to the planet, on the contrary, their equilibrium orbital plane coincides with the equator of the planet and precesses with it as a whole \citep{Goldreich_1965}; therefore, close-in satellites artificially increase the oblateness and the rotational angular momentum of the planet. In the close-in satellite regime, an expression for the effective precession constant has been derived by \cite{Ward_1975}. As detailed by \cite{French-etal_1993}, it consists in replacing $J_2$ and $\lambda$ in Eq.~\eqref{eq:alpha} by the effective values:
   \begin{equation}\label{eq:J2prime}
      J_2' = J_2 + \frac{1}{2}\sum_k\frac{m_k}{M}\frac{a_k^2}{R_\mathrm{eq}^2}
      \quad\text{and}\quad
      \lambda' = \lambda + \sum_k\frac{m_k}{M}\frac{a_k^2}{R_\mathrm{eq}^2}\frac{n_k}{\omega}\,,
   \end{equation}
   where $m_k$, $a_k$, and $n_k$ are the mass, the semi-major axis, and the mean motion of the $k$th satellite. In these expressions, the eccentricities and inclinations of the satellites are neglected. This approximation has been widely used in the literature. In the case of a single satellite, \cite{Boue-Laskar_2006} have obtained a general expression for the precession rate of a planet with an eccentric and inclined satellite, encompassing both the close-in and far-away regimes. Using their article, we can verify that the Galilean satellites are in the close-in regime. The Laplace plane of Callisto is inclined today by less than $1^\circ$ with respect to Jupiter's equator. The small eccentricities and inclinations of the Galilean satellites would contribute to $J_2'$ and $\lambda'$ with terms of order $e_k^2$ and $\eta_k^2$, so even if $e_k$ increases up to $0.1$ (a value found by \citealp{Lari-etal_2020} in some cases) or if $I_k$ increases up to $10^\circ$, the additional contribution to $J_2'$ and $\lambda'$ would only be of order $10^{-4}$ and $10^{-6}$, respectively. As we see below, this contribution is much smaller than our uncertainty on the value of $\lambda$, allowing us to stick to the approximation given by Eq.~\eqref{eq:J2prime}.

   \subsection{Orbital solution}\label{ssec:orbitsol}
   The Hamiltonian given in Eq.~\eqref{eq:Hinit} depends on the orbit of the planet and on its temporal variations. In order to explore the long-term dynamics of Jupiter's spin axis, we need an orbital solution that is valid over billions of years. This is well beyond the timespan covered by ephemerides. Luckily, the orbital dynamics of the giant planets of the Solar System are almost integrable and excellent solutions have been developed. We use the secular solution of \cite{Laskar_1990} expanded in quasi-periodic series:
   \begin{equation}\label{eq:qprep}
      \begin{aligned}
         z = e\exp(i\varpi) &= \sum_k E_k\exp(i\theta_k) \,,\\
         \zeta = \eta\exp(i\Omega) &= \sum_k S_k\exp(i\phi_k)\,,
      \end{aligned}
   \end{equation}
   where $\varpi$ is Jupiter's longitude of perihelion. The amplitudes $E_k$ and $S_k$ are real constants, and the angles $\theta_k$ and $\phi_k$ evolve linearly over time $t$, with frequencies $\mu_k$ and $\nu_k$:
   \begin{equation}\label{eq:munu}
      \theta_k(t) = \mu_k\,t + \theta_k^{(0)}
      \hspace{0.5cm}\text{and}\hspace{0.5cm}
      \phi_k(t) = \nu_k\,t + \phi_k^{(0)}\,.
   \end{equation}
   The complete orbital solution of \cite{Laskar_1990} can be found in Appendix~\ref{asec:QPS} for amplitudes down to $10^{-8}$.
   
   The series in Eq.~\eqref{eq:qprep} contain contributions from all the planets of the Solar System. In the integrable approximation, the frequency of each term corresponds to a unique combination of the fundamental frequencies of the system, usually noted $g_j$ and $s_j$. In the limit of small masses, small eccentricities and small inclinations (Lagrange-Laplace secular system), the $z$ series only contains the frequencies $g_j$, while the $\zeta$ series only contains the frequencies $s_j$ (see e.g. \citealp{Murray-Dermott_1999} or \citealp{Laskar-etal_2012}). This is not the case in more realistic situations, as recalled for instance by \cite{Kreyche-etal_2020} in the context of obliquity dynamics. In planetary systems featuring mean-motion resonances, the spin axis of a planet can be affected by shifted orbital precession frequencies \citep{Millholland-Laughlin_2019} or by secondary resonances \citep{Quillen-etal_2017,Quillen-etal_2018}. However, this does not apply in the Solar System as it is today, even when the existing near commensurabilities (like the ``great Jupiter--Saturn inequality'') are taken into account. Table~\ref{tab:zetashort} shows the combinations of fundamental frequencies identified for the largest terms of Jupiter's $\zeta$ series obtained by \cite{Laskar_1990}.
   
   \begin{table}
      \caption{First twenty terms of Jupiter's inclination and longitude of ascending node in the J2000 equator and equinox reference frame.}
      \label{tab:zetashort}
      \vspace{-0.7cm}
      \begin{equation*}
         \begin{array}{rcrrr}
         \hline
         \hline
         k & \text{identification}\tablefootmark{*} & \nu_k\ (''\cdot\text{yr}^{-1}) & S_k\times 10^8 & \phi_k^{(0)}\ (^\text{o}) \\
         \hline
          1 &          s_5 &   0.00000 & 1377467 & 107.59 \\
          2 &          s_6 & -26.33023 &  315119 & 307.29 \\
          3 &          s_8 &  -0.69189 &   58088 &  23.96 \\
          4 &          s_7 &  -3.00557 &   48134 & 140.33 \\
          5 &  g_5-g_6+s_7 & -26.97744 &    2308 & 222.98 \\
          6 & -g_5+g_6+s_6 &  -2.35835 &    1611 &  44.74 \\
          7 &     2g_6-s_6 &  82.77163 &    1372 & 308.95 \\
          8 &  g_5-g_7+s_7 &  -1.84625 &    1130 &  36.64 \\
          9 &          s_1 &  -5.61755 &    1075 & 168.70 \\
         10 & -g_5+g_7+s_7 &  -4.16482 &     946 &  51.54 \\
         11 &  g_5+g_6-s_6 &  58.80017 &     804 &  32.90 \\
         12 &  g_5-g_6+s_6 & -50.30212 &     691 &  29.84 \\
         13 &     2g_5-s_6 &  34.82788 &     636 & 114.12 \\
         14 &  g_7-g_8+s_7 &  -0.58033 &     565 &  17.32 \\
         15 &          s_2 &  -7.07963 &     454 & 273.79 \\
         16 & -g_5+g_7+s_6 & -27.48935 &     407 &  38.53 \\
         17 &  g_5-g_7+s_6 & -25.17116 &     385 &  35.94 \\
         18 &   s_1+\gamma &  -5.50098 &     383 & 162.89 \\
         19 & -g_7+g_8+s_8 &  -3.11725 &     321 & 326.97 \\
         20 &  s_2+2\gamma &  -6.84091 &     267 & 106.20 \\
         \hline
         \end{array}
      \end{equation*}
      \vspace{-0.5cm}
      \tablefoot{Due to the secular resonance $(g_1-g_5)-(s_1-s_2)$, an additional fundamental frequency $\gamma$ appears in terms 18 and 20 (see \citealp{Laskar_1990}).\\
      \tablefoottext{*}{There is a typographical error in \cite{Laskar_1990} in the identification of the 8th term.}
      }
   \end{table}
   
   As explained by \cite{Saillenfest-etal_2019a}, at first order in the amplitudes $S_k$ and $E_k$, secular spin-orbit resonant angles can only be of the form $\sigma_p = \psi+\phi_p$, where $p$ is a given index in the $\zeta$ series. Resonances featuring terms of the $z$ series only appear at third order and beyond. For the terrestrial planets of the Solar System, the $z$ and $\zeta$ series converge very slowly, which implies that large resonances are very numerous. These resonances overlap massively and produce wide chaotic zones in the obliquity dynamics (see \citealp{Laskar-Robutel_1993,NerondeSurgy-Laskar_1997,Correia-etal_2003,Laskar-etal_2004a}). The situation is very different for the giant planets of the Solar System, for which the $z$ and $\zeta$ series converge quickly owing to the quasi-integrable nature of their dynamics. Therefore, the secular spin-orbit resonances are small and isolated from each other, and only first-order resonances play a substantial role.
   
\begin{figure}
      \centering
      \includegraphics[width=\columnwidth]{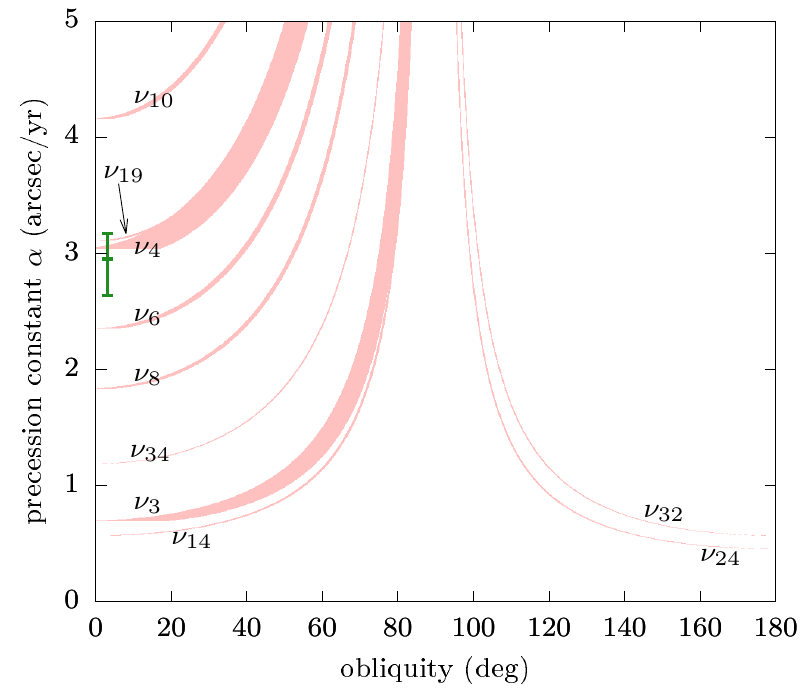}
      \caption{Location and width of every first-order secular spin-orbit resonance for Jupiter. Each resonant angle is of the form $\sigma_p = \psi+\phi_p$ where $\phi_p$ has frequency $\nu_p$ labelled on the graph according to its index in the orbital series (see Table~\ref{tab:zetashort} and Appendix~\ref{asec:QPS}). For a given value of the precession constant $\alpha$, the interval of obliquity enclosed by the separatrix is shown in pink, as computed using the exact formulas given by \cite{Saillenfest-etal_2019a}. The green bar on the left shows Jupiter's current obliquity and the range for its precession constant considered in this article, as detailed in Sects.~\ref{ssec:alpha} and \ref{ssec:condinit}.}
      \label{fig:widths}
   \end{figure}

   Figure~\ref{fig:widths} shows the location and width of every first-order resonance for the spin-axis of Jupiter in an interval of precession constant $\alpha$ ranging from $0''\cdot$yr$^{-1}$ to $5''\cdot$yr$^{-1}$. Because of the chaotic dynamics of the Solar System \citep{Laskar_1989}, the fundamental frequencies related to the terrestrial planets (e.g. $s_1$, $s_2$, and $\gamma$ appearing in Table~\ref{tab:zetashort}) could vary substantially over billions of years \citep{Laskar_1990}. However, they only marginally contribute to Jupiter's orbital solution and none of them takes part in the resonances shown in Fig.~\ref{fig:widths}. Our secular orbital solution of Jupiter can therefore be considered valid over a billion-year timescale.
   
   \subsection{Precession constant}\label{ssec:alpha}
   As shown by the Hamiltonian function in Eq.~\eqref{eq:Hinit}, the precession constant $\alpha$ is a key parameter of the spin-axis dynamics of a planet. Among the physical parameters of Jupiter that enter into its expression (see Eq.~\ref{eq:alpha}), all are very well constrained from observations except the normalised polar moment of inertia $\lambda$.
   
   While comparing the values of $\lambda$ given in the literature, one must be careful about the normalisation used. Equation~\eqref{eq:J2lb} explicitly requires a normalisation using the equatorial radius $R_\text{eq}$, since it is linked to the value of $J_2$. However, published values of the polar moment of inertia are often normalised using the mean radius of Jupiter, which differs from $R_\text{eq}$ by a factor of about $0.978$. This distinction seems to have been missed by \cite{Ward-Canup_2006}, who quote the nominal value given by D.~R.~Williams in the \emph{NASA Jupiter fact sheet}\footnote{\texttt{https://nssdc.gsfc.nasa.gov/planetary/factsheet/\\jupiterfact.html}} as $0.254$, whereas it actually translates into $\lambda=0.243$ when it is normalised using $R_\text{eq}$. \cite{Ward-Canup_2006} also mention that ``theoretical values [of $\lambda$] range from $0.255$ for the extreme of a constant-density core and massless envelope to $0.221$ for a constant-density envelope and point-mass core''. Unfortunately, these numbers are taken from a conference talk given by W.~B. Hubbard in 2005 so we cannot check how they have been obtained. Since Eq.~\eqref{eq:alpha} is used, however, we can assume that they have been properly normalised using $R_\text{eq}$.
   
\begin{figure}
      \centering
      \includegraphics[width=\columnwidth]{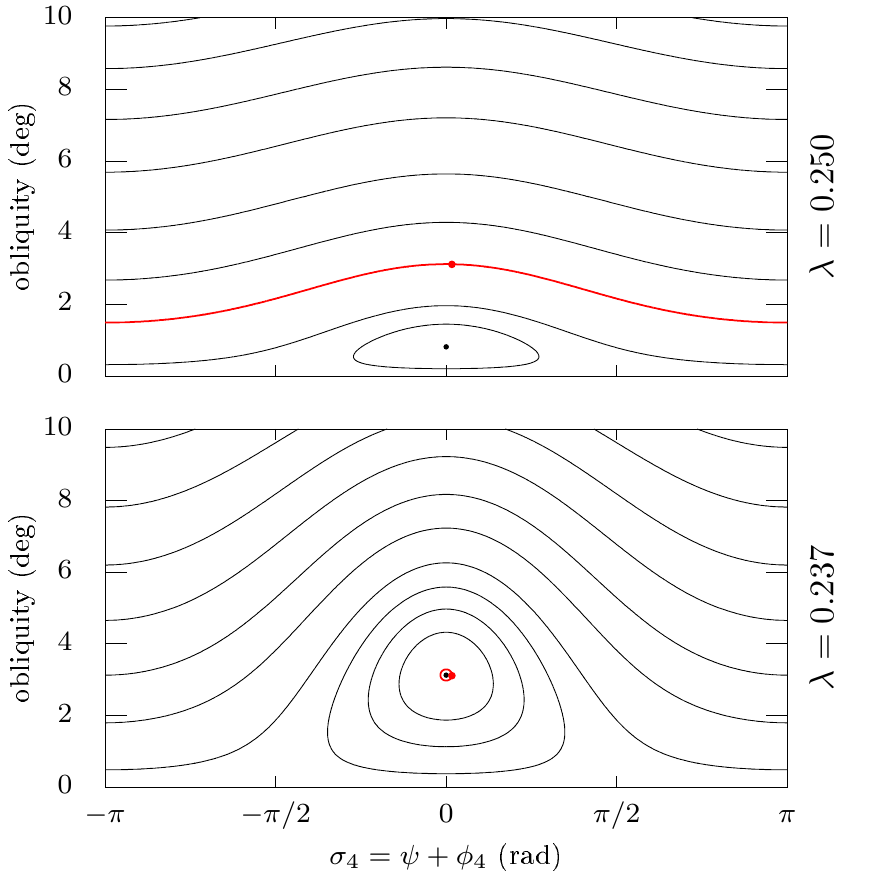}
      \caption{Trajectory of Jupiter's spin axis in the vicinity of resonance with the fourth harmonics of $\zeta$ (see Table~\ref{tab:zetashort}). Being farther away from Jupiter's precession frequency, the contribution of other harmonics can be averaged; their mean contribution is included here up to third order in the amplitudes (as in Eq.~17 of \citealp{Saillenfest-etal_2019a}). Each trajectory corresponds to a level curve of the Hamiltonian, which has only one degree of freedom. The red dot shows the current location of Jupiter, and the black dot shows Cassini state~2. The red curve is the current trajectory of Jupiter's spin axis for $\lambda=0.250$ (top) or $\lambda=0.237$ (bottom).}
      \label{fig:Cassini}
   \end{figure}

   As is shown in Fig.~\ref{fig:widths}, the spin-axis of Jupiter is located very close to a strong secular spin-orbit resonance. The corresponding term of the orbital series is related to the precession mode of Uranus (term $k=4$ in Table~\ref{tab:zetashort}), and the resonant angle is $\sigma_4=\psi+\phi_4$. As noted by \cite{Ward-Canup_2006}, dissipative processes during the early planetary evolution are expected to have forced Jupiter's spin axis to spiral down towards the centre of the resonance, called Cassini state~2. And indeed, the current value of $\sigma_4$ is very close to zero, which has a low probability to happen if Jupiter is far from Cassini state~2 because $\sigma_4$ would then circulate between $0^\circ$ and $360^\circ$. In order to match Cassini state~2, however, Jupiter's normalised moment of inertia should be $\lambda\approx 0.2365$ (see Fig.~\ref{fig:Cassini}). Since this value is not far from what is proposed in the literature, this prompted \cite{Ward-Canup_2006} to consider this value as likely for Jupiter.
   
   As noted by \cite{LeMaistre-etal_2016}, the value of $\lambda\approx 0.2365$ corresponds to a massive core for Jupiter, and estimates obtained from models of Jupiter's interior structure are generally higher. \cite{Helled-etal_2011} obtain values of $\lambda$ ranging from $0.251$ to $0.253$, that were confirmed by \cite{Nettelmann-etal_2012}. These values are consistent with the range of $\lambda\in[0.221,0.255]$ quoted above. Other studies seem to agree on even higher values: \cite{Wahl-etal_2017} and \cite{Ni_2018} present values of $\lambda$ ranging between $0.2629$ and $0.2644$, compatible with the findings of \cite{Hubbard-Marley_1989}, \cite{Nettelmann-etal_2012}, and \cite{Hubbard-Militzer_2016}. Finally, both low and high values are obtained by \cite{Vazan-etal_2016}, who give either $\lambda=0.247$ or $\lambda=0.262$ for three different models. As explained by \cite{LeMaistre-etal_2016}, however, all these values are model-dependent and still a matter of debate. Hopefully, the \emph{Juno} mission will provide direct observational constraints soon that will help us to determine which models of Jupiter's interior structure are the most relevant.
   
   Here, instead of relying on one particular value of $\lambda$, we turn to the exploration of the whole range of values given in the literature, namely $\lambda\in[0.220,0.265]$. The rotation velocity of Jupiter is taken from \cite{Archinal-etal_2018} and the other physical parameters are fixed to those used by \cite{Lari-etal_2020} for consistency with the satellites' orbital evolution (see below). The corresponding value for the current precession constant of Jupiter, computed from Eqs.~\eqref{eq:alpha} and \eqref{eq:J2prime}, ranges from $2.64''\cdot$yr$^{-1}$ to $3.17''\cdot$yr$^{-1}$. Given this large range, using updated physical parameters (see e.g. \citealp{Folkner-etal_2017,Iess-etal_2018,Serra-etal_2019}) would only slightly shift the value of $\alpha$ within our exploration interval.

\begin{figure}
      \centering
      \includegraphics[width=0.8\columnwidth]{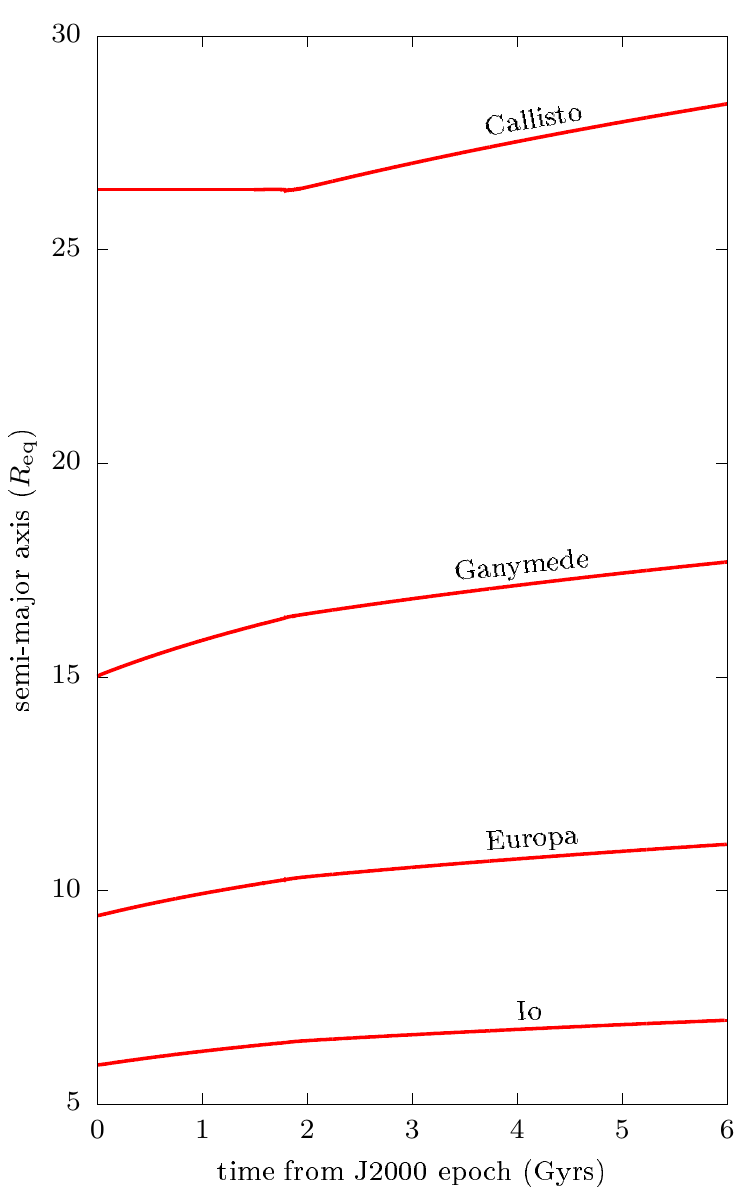}
      \caption{Typical evolution of the semi-major axes of the Galilean satellites obtained by \cite{Lari-etal_2020}. The values are expressed in unit of Jupiter's equatorial radius. The bump at about $1.8$~Gyrs is due to the capture of Callisto into resonance.}
      \label{fig:SatGal2sma}
   \end{figure}

\begin{figure}
      \centering
      \includegraphics[width=\columnwidth]{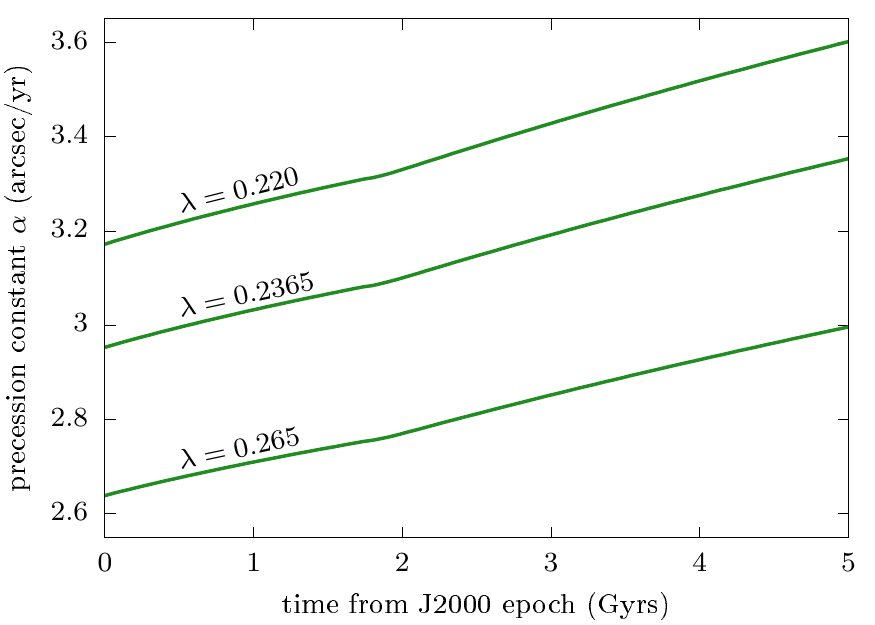}
      \caption{Evolution of the effective precession constant of Jupiter due to the migration of its satellites. The top and bottom curves correspond to the two extreme values of the normalised polar moment of inertia $\lambda$ considered in this article. They appear into $\alpha$ through Eq.~\eqref{eq:alpha}. The central curve corresponds to the value of $\lambda$ that places Jupiter just near Cassini state~2 with the precession mode of Uranus \citep{Ward-Canup_2006}.}
      \label{fig:alphaevol}
   \end{figure}

   Because of tidal dissipation, satellites slowly migrate over time. This produces a drift of the precession constant $\alpha$ on a timescale that is much larger than the precession motion (i.e. the circulation of $\psi$). The long-term spin-axis dynamics of a planet with migrating satellites is described by the Hamiltonian in Eq.~\eqref{eq:Hinit}, but where $\alpha$ is a slowly-varying function of time. In the Earth-Moon system, the outward migration of the Moon produces a decrease of $\alpha$ that pushes the Earth towards a wide chaotic region (see \citealp{NerondeSurgy-Laskar_1997}). This decrease of $\alpha$ is due to the fact that the Moon is in the far-satellite regime (see \citealp{Boue-Laskar_2006}). The Galilean satellites, on the contrary, are in the close-satellite regime, and their outward migration produces an increase of $\alpha$, as shown by Eq.~\eqref{eq:J2prime}. This increase can be quantified using the long-term orbital solution of \cite{Lari-etal_2020} depicted in Fig.~\ref{fig:SatGal2sma} and interpolating between data points. The result is presented in Fig.~\ref{fig:alphaevol} for the two extreme values of $\lambda$ considered in this article, as well as for the value of $\lambda\approx 0.2365$ proposed by \cite{Ward-Canup_2006}. Despite the various outcomes of the dynamics described by \cite{Lari-etal_2020}, the result on the evolution of $\alpha$ is almost undistinguishable from one of their simulations to another, even if the eccentricities of the satellites are taken into account in Eq.~\eqref{eq:J2prime}. Indeed, the variation of $\alpha$ mostly depends on the drift of the satellites' semi-major axes, which is almost identical in every simulation of \cite{Lari-etal_2020}.
   
   Since the rate of energy dissipation between Jupiter and its satellites is not well known today, the timescale of the drift shown in Figs.~\ref{fig:SatGal2sma} and \ref{fig:alphaevol} could somewhat contract or expand. This point is further discussed in Sect.~\ref{sec:evol}. Moreover, other parameters in Eq.~\eqref{eq:alpha} probably slightly vary over billions of years, such as the spin velocity of Jupiter or its oblateness. We consider that the impact of their variations on the value of $\alpha$ is small and contained within our exploration range.

   \subsection{Initial conditions}\label{ssec:condinit}
   The initial orientation of the spin axis is taken from the solution of \cite{Archinal-etal_2018} averaged over short-period terms. At the level of precision required by our exploratory study, the refined orientation obtained by \cite{Durante-etal_2020} is undistinguishable from this nominal orientation. With respect to Jupiter's secular orbital solution (see Sect.~\ref{ssec:orbitsol}), this gives an obliquity $\varepsilon = 3.120^\circ$ and a precession angle $\psi=-137.223^\circ$ at time J2000. The uncertainty on these values is extremely small compared to the range of $\alpha$ considered (see Sect.~\ref{ssec:alpha}). Since the uncertainty is smaller than the curve width of our figures, we do not consider any error bar on the initial value of $\varepsilon$ and $\psi$.
   
\section{Obliquity evolution with migrating satellites}\label{sec:evol}
   
\begin{figure*}
      \centering
      \includegraphics[width=\textwidth]{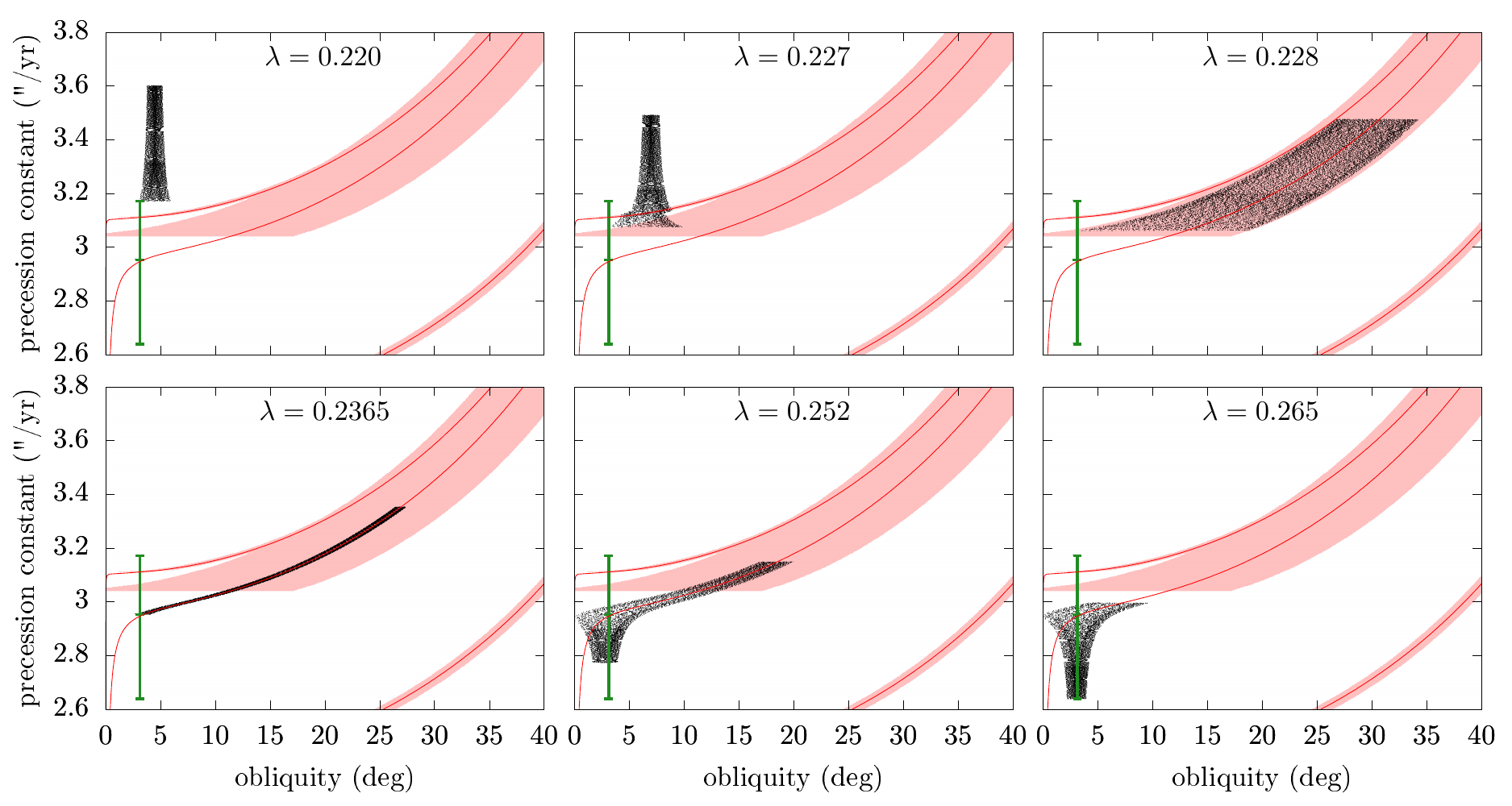}
      \caption{Future evolution of Jupiter's spin axis projected in the plane of the obliquity and the precession constant $\alpha$. Each panel corresponds to a value of the normalised polar moment of inertia of Jupiter $\lambda = C/(MR_\text{eq}^2)$ given in title. The green bar shows the initial location of Jupiter's spin axis according to our exploration interval of $\lambda$; the central mark is the value proposed by \cite{Ward-Canup_2006}. The red curves show the centre of all first-order secular spin-orbit resonances (Cassini state~2) and the coloured areas represent their widths (same as Fig.~\ref{fig:widths}). From bottom to top, the resonances are with $\phi_6$, with $\phi_4$, and with $\phi_{19}$ (see Table~\ref{tab:zetashort}). The black dots show the numerical solutions obtained over a timespan of $5$~Gyrs from now; they evolve from bottom to top. According to the exact migration rate of the Galilean satellites, the timeline could somewhat contract or expand (see text).}
      \label{fig:obevol}
   \end{figure*}

   For values of $\lambda$ finely sampled in our exploration interval, the spin axis of Jupiter is numerically propagated forwards in time for $5$~Gyrs. By virtue of trigonometric identities, moving Jupiter's orbit one step forwards in time using the quasi-periodic decomposition in Eq.~\eqref{eq:qprep} only amounts to computing a few sums and products. The trajectories obtained are shown in Fig.~\ref{fig:obevol} for a few values of $\lambda$. They are projected in the plane of the obliquity and the precession constant of Jupiter, where we localise also the centres and widths of all first-order secular spin-orbit resonances. See Appendix~\ref{asec:phi4} for further details about the geometry of the resonances.
   
\begin{figure}
      \centering
      \includegraphics[width=\columnwidth]{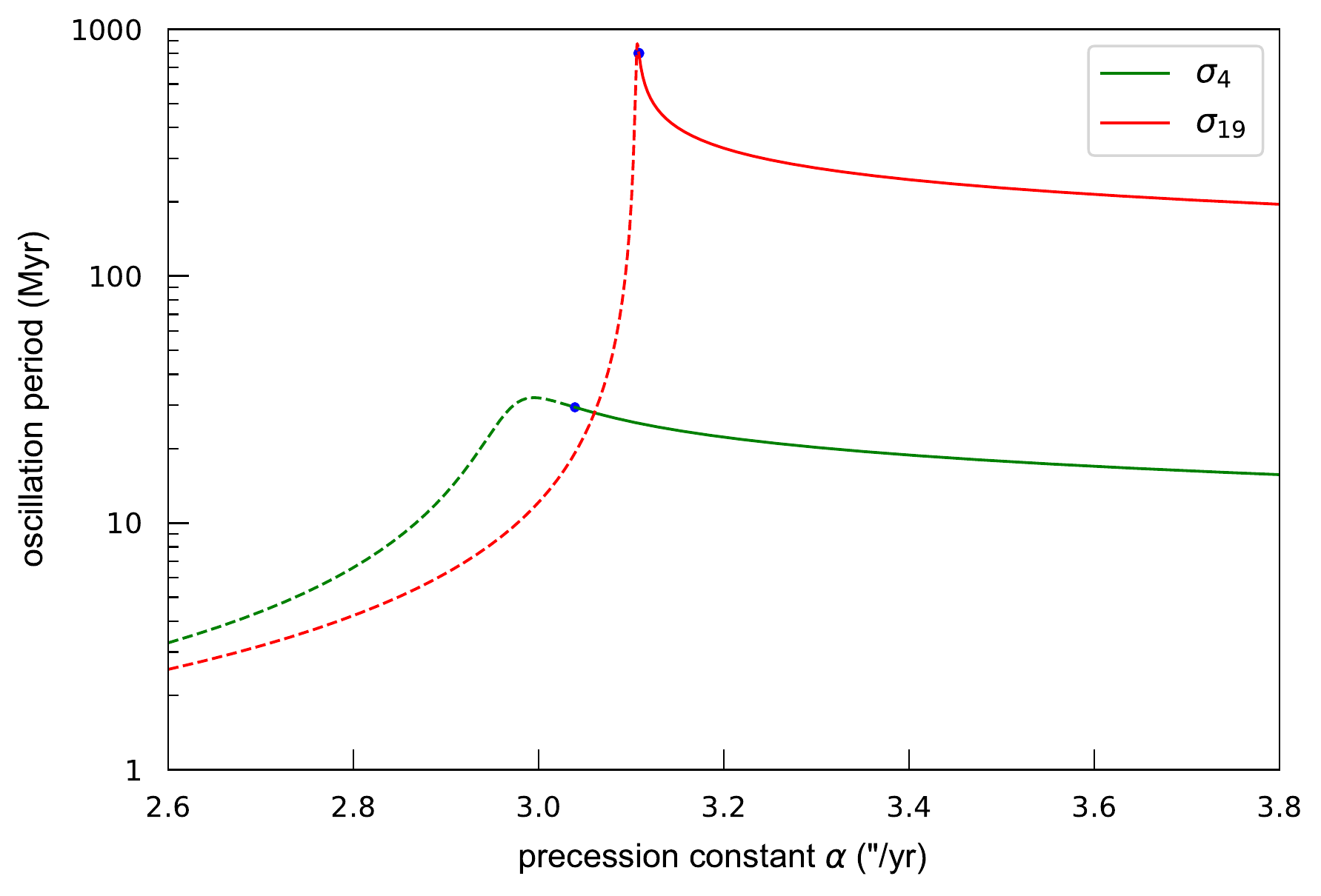}
      \caption{Period of small oscillations about the resonance centre for a resonance with $\phi_4$ or $\phi_{19}$. Even though complete closed-form solutions exist (see \citealp{Haponiak-etal_2020}), the small-oscillation limit leads to handier formulas, suitable for order-of-magnitude estimates. The resonant angles are $\sigma_4=\psi+\phi_4$ and $\sigma_{19}=\psi+\phi_{19}$, respectively. Dashed curves are used for oscillations about Cassini state 2 before the separatrix appears. The appearance of the separatrix is marked by a blue dot.}
      \label{fig:Plib}
   \end{figure}

   For values of $\lambda$ smaller than about $0.228$, Jupiter starts outside of the large resonance with $\phi_4$, and the increase of its precession constant $\alpha$ pushes it even farther away over time. As shown by the trajectory computed for $\lambda=0.227$, the crossing of the very thin resonance with $\phi_{19}$ twists the trajectory a little, but this cannot produce any large change of obliquity. Indeed the resonance with $\phi_{19}$ is not strong enough to capture Jupiter's spin axis: It is crossed quickly as $\alpha$ increases, and Fig.~\ref{fig:Plib} shows that the libration period of $\sigma_{19} = \psi+\phi_{19}$ is very long. This results in a non-adiabatic crossing (see Appendix~\ref{asec:phi19} for details). Consequently, no major obliquity variation for Jupiter can be expected in the future if $\lambda<0.228$. However, such small values of $\lambda$ seem to be ruled out by most models of Jupiter's interior (see Sect.~\ref{ssec:alpha}).
   
   For values of $\lambda$ larger than $0.228$, on the contrary, Jupiter is currently located inside or below the large resonance with $\phi_4$. As predicted, the value $\lambda=0.2365$ results in very small oscillations around Cassini state~2. As its precession constant $\alpha$ slowly increases with time, Jupiter is captured into the resonance and follows the drift of its centre towards large obliquities. Indeed, the resonance with $\phi_4$ is large, and the libration period of $\sigma_4=\psi+\phi_4$ is short compared to the variation timescale of $\alpha$ (see Figs.~\ref{fig:obevol} and \ref{fig:Plib}). This results in an adiabatic capture. The various possible outcomes of adiabatic and non-adiabatic crossings of secular spin-orbit resonances have recently been studied by \cite{Su-etal_2020}. However, the orbital motion is here not limited to a single harmonic, and Appendix~\ref{asec:phi19} shows that the separatrix of the resonance is replaced by a chaotic ``moat''. Properly speaking, the resonance with $\phi_4$ becomes a ``true resonance'' only as soon as the separatrix appears, that is, for $\alpha$ larger than $\alpha\approx 3.04''\cdot$yr$^{-1}$ (see Appendix~\ref{asec:phi4}). In the whole range of values of $\lambda>0.228$ considered in this article, the spin axis of Jupiter is initially located close enough to Cassini state~2 to invariably end up inside the separatrix of the resonance when it appears. The capture probability is therefore $100\%$. None of our simulation shows a release out of resonance or a turn-off towards Cassini state~1, which could have been a possible outcome if Jupiter's spin axis was initially located farther away from Cassini state~2 or if the drift of $\alpha$ was not adiabatic\footnote{There is a typographical error in \cite{Saillenfest-etal_2019a}: the list of the Cassini states given before Eq.~(22) should read (4,2,3,1) instead of (1,2,3,4) in order to match the denomination introduced by \cite{Peale_1969}.}. Since in canonical coordinates the resonance width increases for $\alpha$ growing up to $4.244''\cdot$yr$^{-1}$, no separatrix crossing can happen, even for a large libration amplitude inside the resonance  (e.g. for $\lambda=0.228$ in Fig.~\ref{fig:obevol}). The maximum obliquity reached by Jupiter is therefore only limited by the finite amount of time considered.
   
\begin{figure}
      \centering
      \includegraphics[width=\columnwidth]{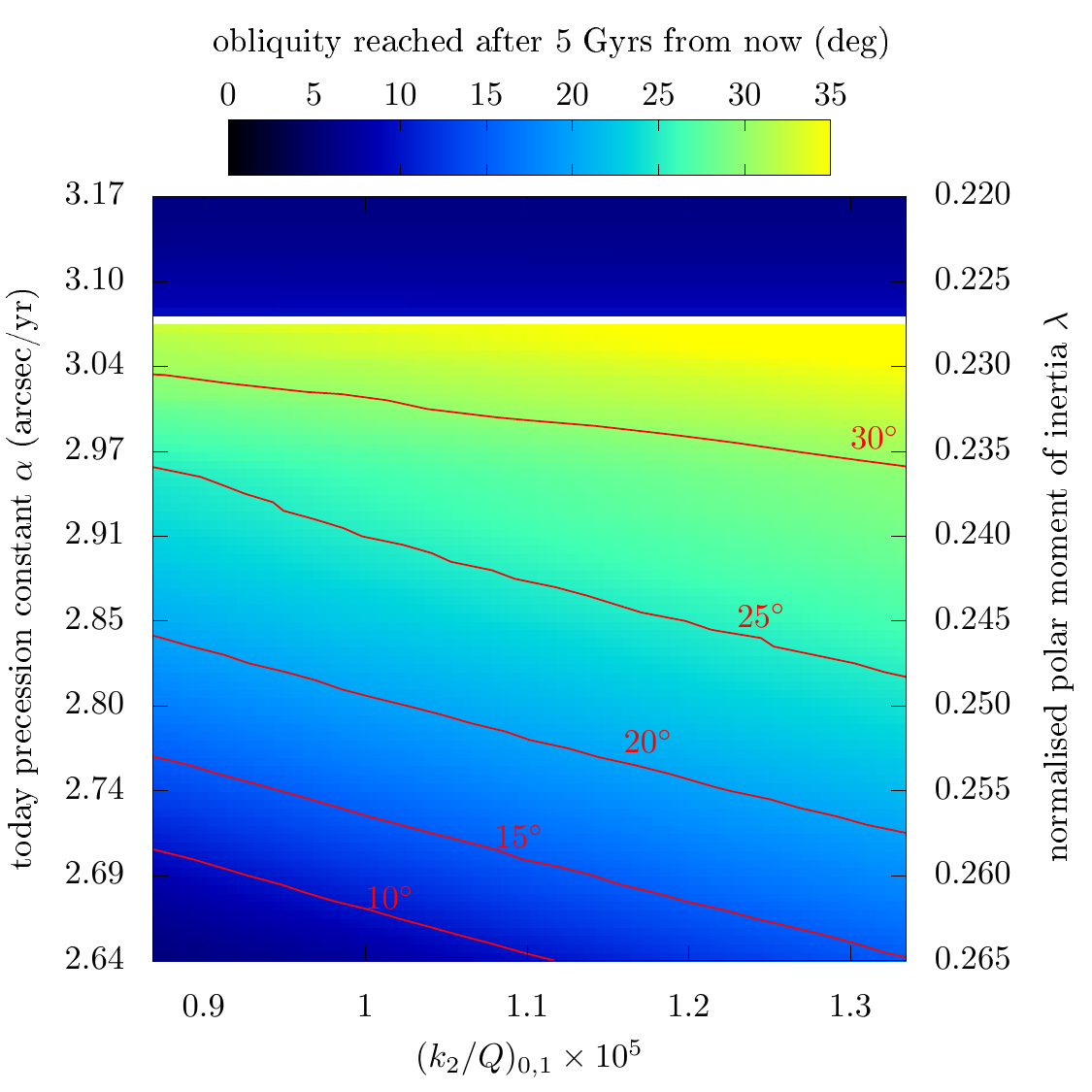}
      \caption{Maximum obliquity reached by Jupiter after $5$~Gyrs from now as a function of its normalised polar moment of inertia $\lambda = C/(MR_\text{eq}^2)$ (right vertical axis) and the dissipation parameter of Jupiter at Io's frequency (horizontal axis). The left vertical axis shows the current precession constant $\alpha$ of Jupiter. Some level curves are shown in red.}
      \label{fig:obmax}
   \end{figure}

   If the Galilean satellites migrate faster than shown in Fig.~\ref{fig:SatGal2sma}, the obliquity reached in $5$~Gyrs would be larger than that presented in Fig.~\ref{fig:obevol}. The migration rate of the satellites is not well known. According to \cite{Lari-etal_2020}, the long-term migration rate of the satellites varies by $\pm 15\%$ over the uncertainty range of the parameter $(k_2/Q)_{0,1}$ measured by \cite{Lainey-etal_2009}. This parameter quantifies the dissipation within Jupiter at Io's frequency. Figure~\ref{fig:obmax} shows the maximum obliquity reached in $5$~Gyrs for $\lambda$ sampled in our exploration interval and $(k_2/Q)_{0,1}$ sampled in its uncertainty range. We retrieve the discontinuity at $\lambda\approx 0.228$ discussed before, below which only small obliquity variations are possible. For $\lambda > 0.228$, as expected, we see that a fast migration and a small moment of inertia produce a fast increase of obliquity, which reaches $37^\circ$ in $5$~Gyrs in the most favourable case of Fig.~\ref{fig:obmax}. On the contrary, a slow migration and a large moment of inertia produce a slow increase of obliquity, which barely reaches $6^\circ$ in $5$~Gyrs in the most unfavourable case of Fig.~\ref{fig:obmax}.
   
\section{Discussion and conclusion}\label{sec:ccl}
   Prompted by the peculiarly small value of the current obliquity of Jupiter, we studied the future long-term evolution of its spin axis under the influence of its slowly migrating satellites.
   
   Jupiter is located today near a strong secular spin-orbit resonance with the nodal precession mode of Uranus \citep{Ward-Canup_2006}. Because of this resonance, the obliquity of Jupiter is found to be currently increasing, provided that its normalised polar moment of inertia $\lambda = C/(MR_\text{eq}^2)$ is larger than about $0.228$. Such a small value seems to be ruled out by models of Jupiter's interior (see e.g. \citealp{Helled-etal_2011,Hubbard-Militzer_2016,Wahl-etal_2017}). For larger values of $\lambda$, the migration of the Galilean satellites induces an adiabatic drift of the precession constant $\alpha$ of Jupiter that pushes its spin axis inside the resonance and forces it to follow the resonance centre towards high obliquities. For the value $\lambda\approx 0.2365$ proposed by \cite{Ward-Canup_2006}, the obliquity can reach values as large as $30^\circ$ in the next $5$~Gyrs. For the value $\lambda\approx 0.252$ obtained by \cite{Helled-etal_2011}, the obliquity reaches values ranging from about $17^\circ$ to $23^\circ$. The increase is more modest for values close to $\lambda\approx 0.264$ found by other authors, for which the maximum value of the obliquity ranges from about $6^\circ$ to $17^\circ$. Hence, our main conclusion is that, contrary to Saturn, Jupiter did not have time to tilt much yet from its primordial orientation, but it will in the future and possibly a lot.
   
   The model of tidal dissipation applied by \cite{Lari-etal_2020} to the Galilean satellites and used here to compute the drift of $\alpha$ is simplified. The current migration rates of satellites in the Solar System have been proved to be higher than previously thought (see \citealp{Lainey-etal_2009,Lainey-etal_2017}). As discussed by \cite{Lari-etal_2020}, the migration of the Galilean satellites could be even faster than considered here if ever one of the outer satellites was pushed by a resonance with the frequency of an internal oscillation of Jupiter \citep{Fuller-etal_2016}. This would result in a faster increase of Jupiter's obliquity. This increase would be halted, however, if the satellites ever migrate so fast as to break the adiabaticity of the capture into secular spin-orbit resonance. In this case, Jupiter would cross the resonance and exit without following the drift of its centre (see e.g. \citealp{Ward-Hamilton_2004,Su-etal_2020}). Numerical experiments show that adiabaticity would be broken for a migration more than $110$~times faster than currently estimated. Such an extremely fast migration seems unlikely. Moreover, with such a fast migration, Callisto and then Ganymede would soon go beyond the close-satellite regime \citep{Boue-Laskar_2006}: This would slow down the increase of $\alpha$ and possibly restore the adiabaticity of its drift. Therefore, the future increase of Jupiter's obliquity appears to be a robust result.
   
   The maximum obliquity that Jupiter will reach could be very large, but it depends on the precise value of Jupiter's polar moment of inertia and on the precise migration rate of the Galilean satellites. We hope to obtain soon new estimates for these two crucial parameters, in particular from the results of the \emph{Juno} and JUICE missions.
    
   
\begin{acknowledgements}
   We thank Marco Fenucci for his help and his suggestions during the redaction of our manuscript. We also thank the anonymous referee for her/his valuable comments. G.~L. acknowledges financial support from the Italian Space Agency (ASI) through agreement 2017-40-H.0 in the context of the NASA \emph{Juno} mission.
\end{acknowledgements}

\bibliographystyle{aa}
\bibliography{satgal2Jupiter}

\appendix
   
\section{Orbital solution for Jupiter}\label{asec:QPS}
The secular orbital solution of \cite{Laskar_1990} is obtained by multiplying the normalised proper modes $z_i^\bullet$ and $\zeta_i^\bullet$ (Tables VI and VII of \citealp{Laskar_1990}) by  the matrix $\tilde{S}$ corresponding to the linear part of the solution (Table V of \citealp{Laskar_1990}). In the series obtained, the terms with the same combination of frequencies are then merged together, resulting in 56 terms in eccentricity and 60 terms in inclination. This forms the secular part of the orbital solution of Jupiter, which is what is required by our averaged model.

The orbital solution is expressed in the variables $z$ and $\zeta$ as described in Eqs.~\eqref{eq:qprep} and \eqref{eq:munu}. In Tables~\ref{tab:z} and \ref{tab:zeta}, we give the terms of the solution in the J2000 ecliptic and equinox reference frame for amplitudes down to $10^{-8}$.

\begin{table}
   \caption{Quasi-periodic decomposition of Jupiter's eccentricity and longitude of perihelion (variable $z$).}
   \label{tab:z}
   \vspace{-0.7cm}
   \small
   \begin{equation*}
      \begin{array}{rrrr}
      \hline
      \hline
      k & \mu_k\ (''\cdot\text{yr}^{-1}) & E_k\times 10^8 & \theta_k^{(0)}\ (^\text{o}) \\
      \hline
       1 &   4.24882 & 4411915 &  30.67 \\
       2 &  28.22069 & 1574994 & 308.11 \\
       3 &   3.08952 &  180018 & 121.36 \\
       4 &  52.19257 &   51596 &  45.55 \\
       5 &  27.06140 &   18405 & 218.71 \\
       6 &  29.37998 &   17762 & 217.54 \\
       7 &  28.86795 &   10743 &  32.64 \\
       8 &  27.57346 &    9436 &  43.74 \\
       9 &   5.40817 &    6135 & 120.31 \\
      10 &   0.66708 &    5755 &  73.98 \\
      11 &  53.35188 &    4415 & 314.90 \\
      12 &  76.16447 &    2441 & 143.03 \\
      13 &  51.03334 &    1354 & 316.29 \\
      14 &   7.45592 &    1354 &  20.24 \\
      15 & -19.72306 &    1083 & 293.24 \\
      16 &   4.89647 &     982 & 291.61 \\
      17 &   5.59644 &     941 & 290.35 \\
      18 &   1.93168 &     767 & 198.10 \\
      19 &   3.60029 &     543 & 121.39 \\
      20 & -56.90922 &     485 &  44.11 \\
      21 &   2.97706 &     470 & 306.81 \\
      22 &   5.47449 &     354 &  95.01 \\
      23 &  17.91550 &     295 & 155.35 \\
      24 &   5.71670 &     269 & 300.52 \\
      25 & -20.88236 &     222 & 203.93 \\
      26 &   6.93423 &     173 & 349.25 \\
      27 &   1.82121 &     161 & 150.50 \\
      28 &   5.35823 &     145 & 274.88 \\
      29 &   7.05595 &     139 & 178.82 \\
      30 &   7.34103 &     114 &  27.85 \\
      31 &  17.36469 &     101 & 123.95 \\
      32 &   0.77840 &      99 &  65.10 \\
      33 &   7.57299 &      80 & 191.47 \\
      34 &   5.99227 &      53 & 293.56 \\
      35 &   5.65485 &      51 & 219.22 \\
      36 &   4.36906 &      49 &  40.82 \\
      37 &   5.23841 &      43 &  92.97 \\
      38 &   6.82468 &      37 &  14.53 \\
      39 &  -0.49216 &      29 & 164.74 \\
      40 &  17.08266 &      28 & 179.38 \\
      41 &  16.81285 &      27 & 273.37 \\
      42 &   7.20563 &      23 & 323.91 \\
      43 &   7.71663 &      15 & 273.52 \\
      44 &  19.01870 &      10 & 219.75 \\
      45 &  17.15752 &      10 & 325.02 \\
      46 &  16.52731 &       6 & 131.91 \\
      47 &  17.63081 &       6 & 183.87 \\
      48 &  17.81084 &       6 &  58.56 \\
      49 &  18.18553 &       5 &  57.27 \\
      50 &  17.47683 &       5 & 260.26 \\
      51 &  17.72293 &       4 &  48.46 \\
      52 &  17.55234 &       4 & 197.65 \\
      53 &  18.01611 &       4 &  44.83 \\
      54 &  16.26122 &       2 &  58.89 \\
      55 &  18.08627 &       2 & 356.17 \\
      56 &  18.46794 &       1 & 209.01 \\
      \hline
      \end{array}
   \end{equation*}
   \vspace{-0.5cm}
   \tablefoot{This solution has been directly obtained from \cite{Laskar_1990} as explained in the text. The phases $\theta_k^{(0)}$ are given at time J2000.}
\end{table}

\begin{table}
   \caption{Quasi-periodic decomposition of Jupiter's inclination and longitude of ascending node (variable $\zeta$).}
   \label{tab:zeta}
   \vspace{-0.7cm}
   \small
   \begin{equation*}
      \begin{array}{rrrr}
      \hline
      \hline
      k & \nu_k\ (''\cdot\text{yr}^{-1}) & S_k\times 10^8 & \phi_k^{(0)}\ (^\text{o}) \\
      \hline
       1 &   0.00000 & 1377467 & 107.59 \\
       2 & -26.33023 &  315119 & 307.29 \\
       3 &  -0.69189 &   58088 &  23.96 \\
       4 &  -3.00557 &   48134 & 140.33 \\
       5 & -26.97744 &    2308 & 222.98 \\
       6 &  -2.35835 &    1611 &  44.74 \\
       7 &  82.77163 &    1372 & 308.95 \\
       8 &  -1.84625 &    1130 &  36.64 \\
       9 &  -5.61755 &    1075 & 168.70 \\
      10 &  -4.16482 &     946 &  51.54 \\
      11 &  58.80017 &     804 &  32.90 \\
      12 & -50.30212 &     691 &  29.84 \\
      13 &  34.82788 &     636 & 114.12 \\
      14 &  -0.58033 &     565 &  17.32 \\
      15 &  -7.07963 &     454 & 273.79 \\
      16 & -27.48935 &     407 &  38.53 \\
      17 & -25.17116 &     385 &  35.94 \\
      18 &  -5.50098 &     383 & 162.89 \\
      19 &  -3.11725 &     321 & 326.97 \\
      20 &  -6.84091 &     267 & 106.20 \\
      21 & -28.13656 &     256 & 134.07 \\
      22 &  -7.19493 &     226 & 105.14 \\
      23 &  -6.96094 &     215 &  97.96 \\
      24 &   0.46547 &     162 & 286.88 \\
      25 & -17.74818 &     149 & 123.28 \\
      26 &  -7.33264 &     144 & 196.75 \\
      27 &  -5.85017 &     130 & 345.47 \\
      28 &  11.50319 &     103 & 281.01 \\
      29 &  -5.21610 &      97 & 198.91 \\
      30 &  -5.37178 &      97 & 215.48 \\
      31 &  -5.10025 &      94 &  15.38 \\
      32 &   0.57829 &      56 & 103.72 \\
      33 &  -5.96899 &      55 & 170.64 \\
      34 &  -1.19906 &      53 & 133.26 \\
      35 &  -6.73842 &      47 &  44.50 \\
      36 &  -7.40536 &      44 & 233.35 \\
      37 &  -7.48780 &      40 &  47.95 \\
      38 &  -6.15490 &      40 & 269.77 \\
      39 &  20.96631 &      40 &  57.78 \\
      40 &  -6.56016 &      38 & 303.47 \\
      41 &   9.18847 &      32 &   1.15 \\
      42 &  -8.42342 &      32 & 211.21 \\
      43 &  10.34389 &      32 & 190.85 \\
      44 & -18.85115 &      23 & 240.06 \\
      45 & -17.19656 &      17 & 334.19 \\
      46 &  18.14984 &      15 & 291.19 \\
      47 & -19.40256 &      14 & 207.96 \\
      48 & -18.01114 &      11 & 242.09 \\
      49 & -17.66094 &      11 & 138.93 \\
      50 & -17.83857 &       9 & 289.13 \\
      51 & -17.54636 &       8 & 246.71 \\
      52 & -18.30007 &       6 & 267.05 \\
      53 & -17.94404 &       5 & 212.26 \\
      54 & -18.59563 &       5 &  98.11 \\
      55 & -19.13075 &       1 & 305.90 \\
      \hline
      \end{array}
   \end{equation*}
   \vspace{-0.5cm}
   \tablefoot{This solution has been directly obtained from \cite{Laskar_1990} as explained in the text. The phases $\phi_k^{(0)}$ are given at time J2000.}
\end{table}

\section{Crossing the resonance with $\phi_4$}\label{asec:phi4}

\begin{figure*}
      \centering
      \includegraphics[width=\textwidth]{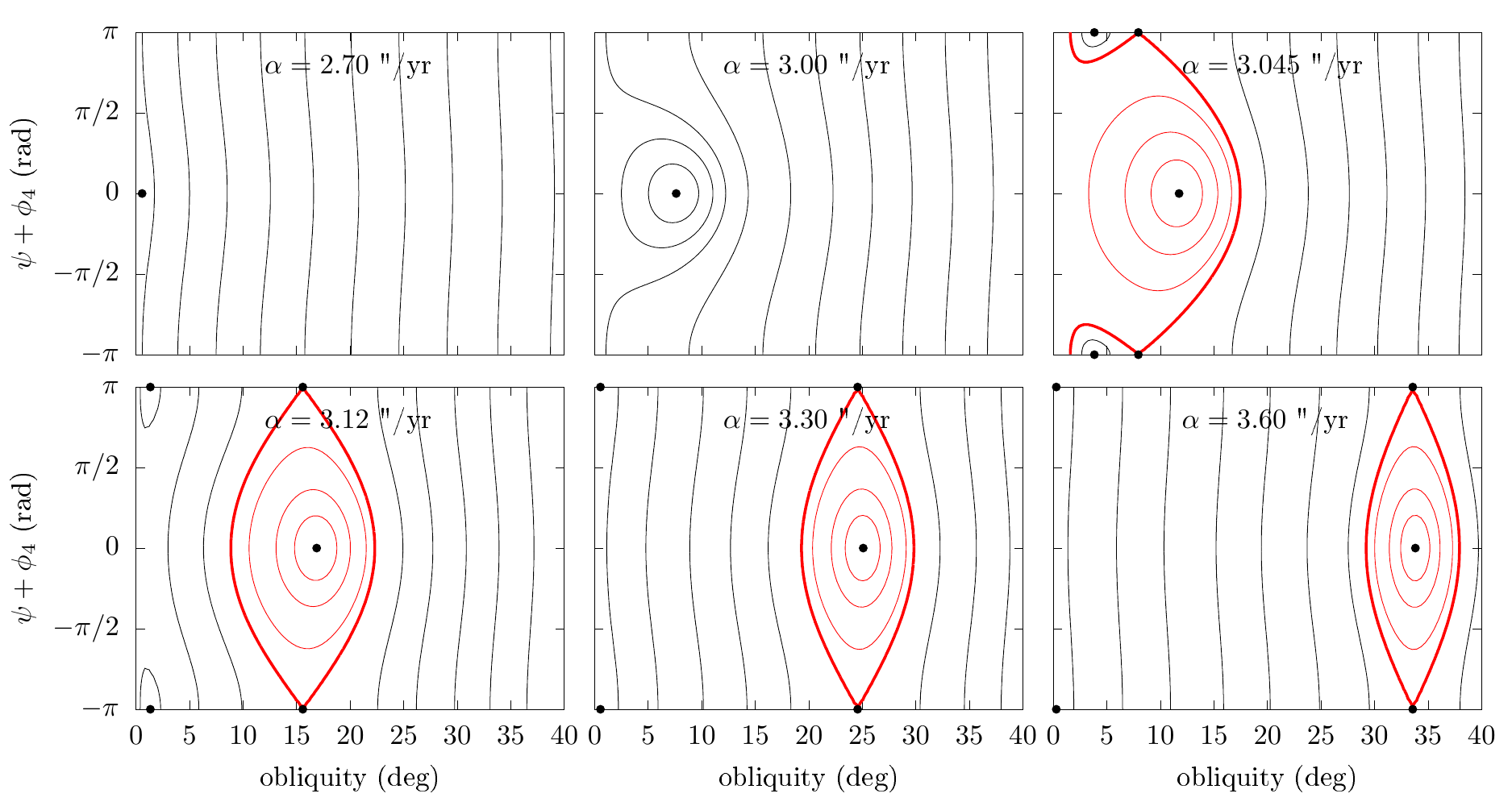}
      \caption{Level curves of the Hamiltonian function in the vicinity of resonance with $\phi_4$. The resonant angle is $\sigma_4=\psi+\phi_4$. Other terms are averaged and included up to the third order of their amplitudes (see \citealp{Saillenfest-etal_2019a}). Each panel corresponds to a different value of the precession constant $\alpha$. Equilibrium points (called Cassini states) are shown by black spots. The interior of the resonance is coloured red and the separatrix is shown with a thick red curve. The location and width of the resonance for continuous values of $\alpha$ can be seen in Fig.~\ref{fig:obevol}. In order to avoid being misled by coordinate singularities, Fig.~\ref{fig:widthscutpolar} shows the same level curves in a different set of coordinates.}
      \label{fig:widthscut}
      \centering
      \includegraphics[width=\textwidth]{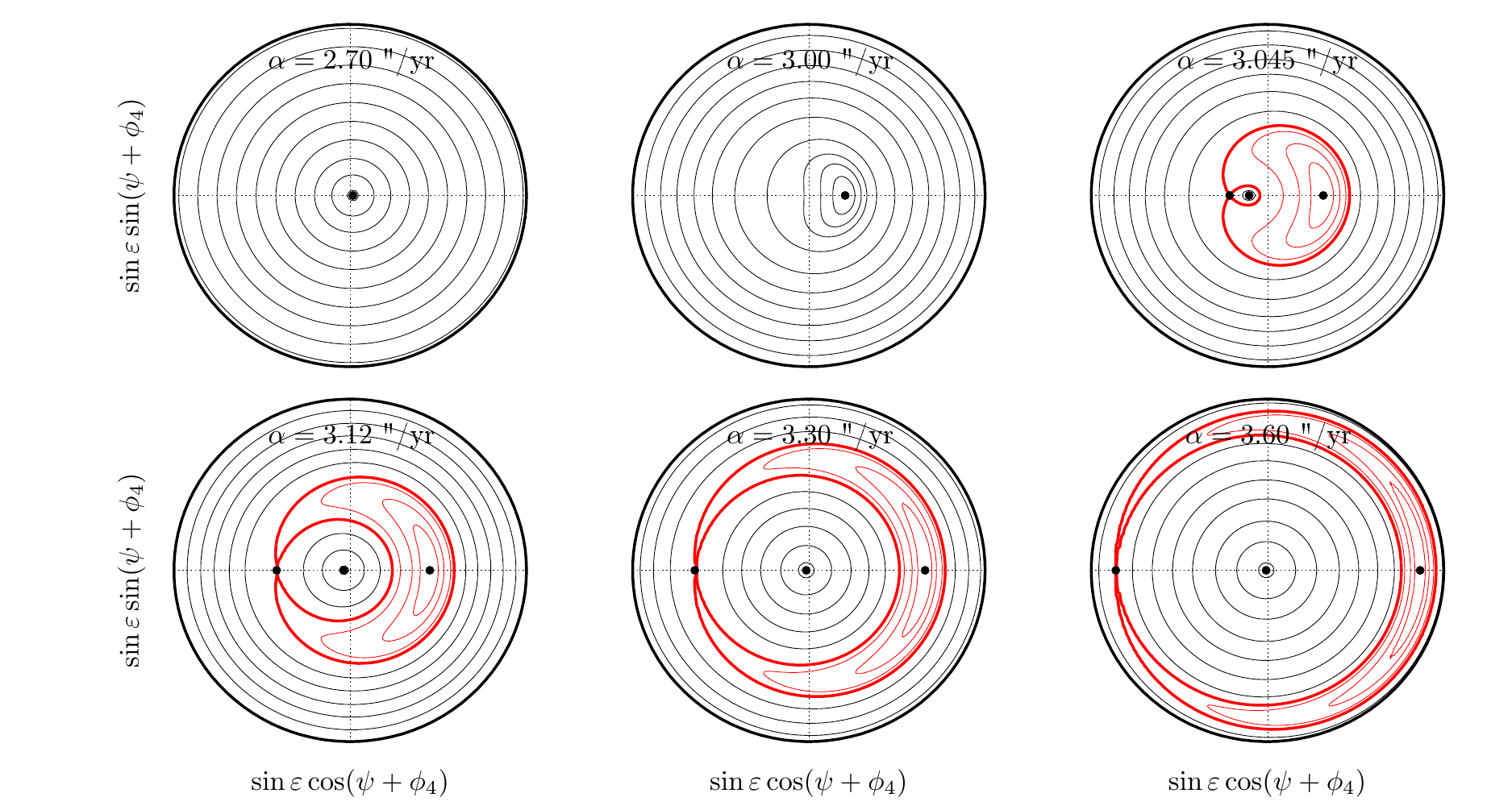}
      \caption{Same as Fig.~\ref{fig:widthscut}, but using polar coordinates that are not singular for an obliquity $\varepsilon=0^\circ$. The outer black circle corresponds to an obliquity $\varepsilon=40^\circ$.}
      \label{fig:widthscutpolar}
   \end{figure*}

  Figures~\ref{fig:widths} and \ref{fig:obevol} show the location and width of all first-order secular spin-orbit resonances produced by Jupiter's orbital solution (Appendix~\ref{asec:QPS}). In particular, Jupiter is located very close to the large resonance with $\phi_4$, whose frequency is the nodal precession mode of Uranus. Figures~\ref{fig:widthscut} and \ref{fig:widthscutpolar} show the geometry of the resonance with $\phi_4$ for different values of the precession constant $\alpha$ of Jupiter. These graphs can be understood as horizontal sections of Fig.~\ref{fig:obevol}, where we can locate the centre of the resonance (i.e. Cassini state~2) and the separatrix width. For easier comparison, Figs.~\ref{fig:obevol} and \ref{fig:widthscut} share the same horizontal axis.
  
\section{Crossing the resonance with $\phi_{19}$}\label{asec:phi19}
   
   As a matter of fact, Jupiter's orbital motion is not restricted to the $\phi_4$ term. However, secular spin-orbit resonances with all other terms (apart from $\phi_{19}$) are located very far from the location of Jupiter, so that their effects average over time. The case of $\phi_{19}$ is special: even though it is very thin, this resonance is not far from Jupiter's location (see the upper red curve in Fig.~\ref{fig:obevol}), which means that $\psi+\phi_{19}$ is a slow angle that cannot be averaged out.
   
\begin{figure*}
      \centering
      \includegraphics[width=0.45\textwidth]{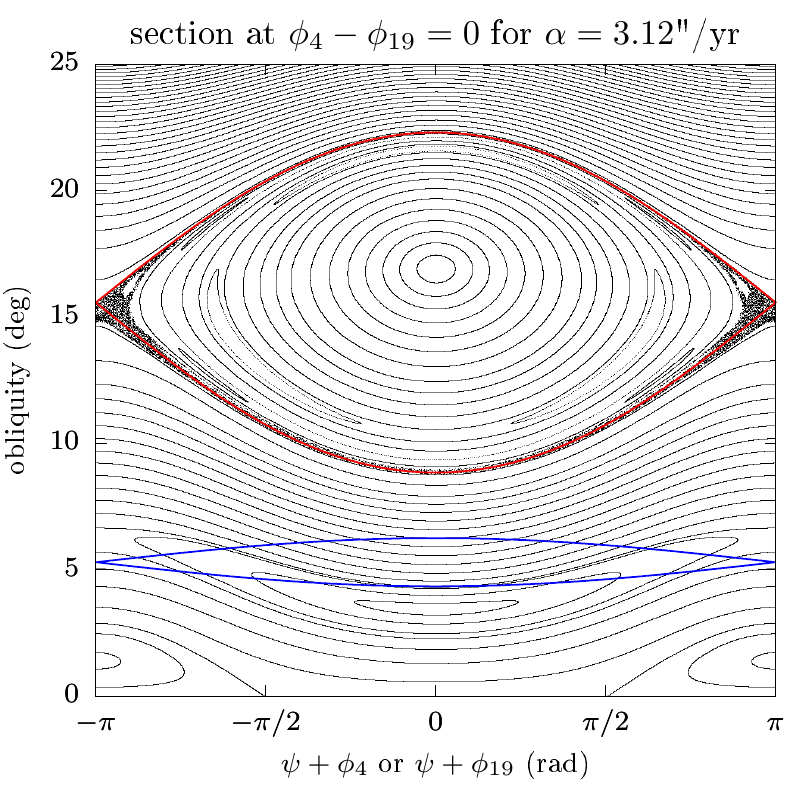}
      \hfill
      \includegraphics[width=0.45\textwidth]{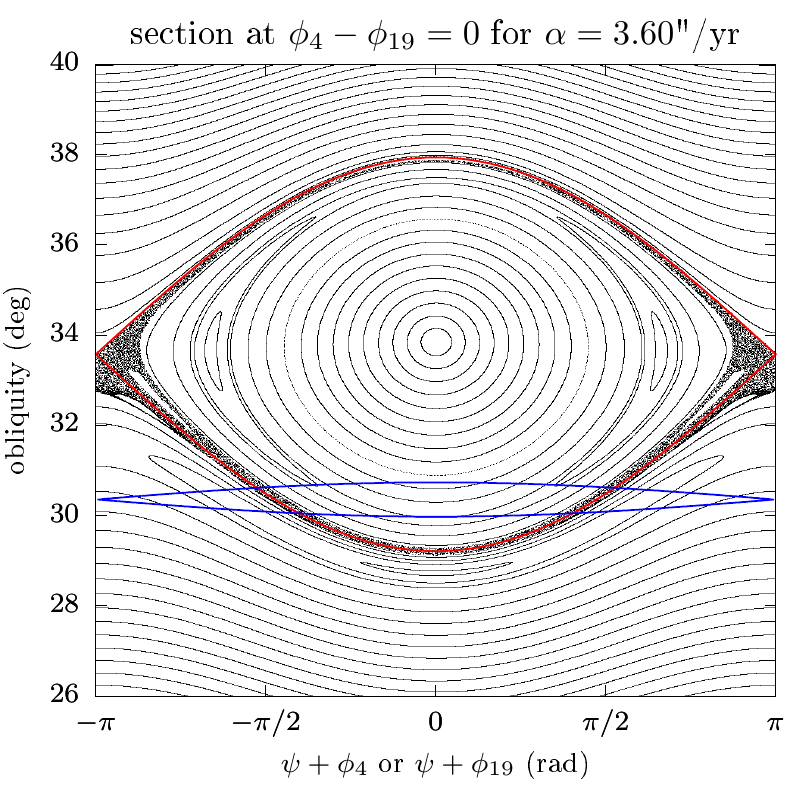}
      \caption{Poincar{\'e} surfaces of section showing the dynamics in the vicinity of resonances with $\phi_4$ and $\phi_{19}$. Each graph corresponds to a different value of $\alpha$ (see titles). The separatrices of the two resonances taken separately are shown with coloured curves: red for~$\phi_4$ and blue for~$\phi_{19}$.}
      \label{fig:PSect}
   \end{figure*}

   Instead of considering only $\phi_4$, as in Appendix~\ref{asec:phi4}, a more rigorous model of the long-term spin-axis dynamics of Jupiter consists in averaging the Hamiltonian function over all angles except resonances with both $\phi_4$ and $\phi_{19}$. For a constant value of $\alpha$, this results in a two-degree-of-freedom Hamiltonian system, in which the two angle coordinates are $\sigma_4=\psi+\phi_4$ and $\sigma_{19}=\psi+\phi_{19}$. The dynamics can then be studied using Poincar{\'e} surfaces of section. Figure~\ref{fig:PSect} shows two examples of sections. The lower island centred at $\sigma_{19}=0$ corresponds to the thin resonance with $\phi_{19}$: As expected, it is completely distorted as compared to the unperturbed separatrix (blue curve) due to the proximity of the large resonance with $\phi_4$. It still persists, however, as a set of periodic orbits. In contrast, the large resonance with $\phi_4$ is hardly affected at all by the $\phi_{19}$ term, which only transforms its separatrix into a thin chaotic belt. In the left panel of Fig.~\ref{fig:PSect}, we can also recognise Cassini state~1 with $\phi_4$ (for $\sigma_4=\pi$ and a small obliquity), that is also visible in Fig.~\ref{fig:widthscut}.
   
   We investigated whether Jupiter could be trapped into the thin resonance with $\phi_{19}$ and follow its resonance centre, but we found out that this can never happen. On the one hand, the current phase of $\sigma_{19}$ is close to $\pi$, so that even if $\lambda$ is finely tuned to place Jupiter right inside the resonance, it ends up near the separatrix, leading to an unstable resonant motion. On the other hand, as shown in Fig.~\ref{fig:Plib}, the libration period of $\sigma_{19}$ is extremely large (the width and oscillation frequency both scale as the square root of the amplitude $S_{19}$). This means that, as $\alpha$ increases, the crossing of this resonance is not adiabatic. The libration periods shown in Fig.~\ref{fig:Plib} should be compared to the time needed for $\alpha$ to go through the resonant region. According to Fig.~\ref{fig:alphaevol}, the mean increase rate of $\alpha$ is $0.086''\cdot$yr$^{-1}\cdot$Gyr$^{-1}$ and according to Fig.~\ref{fig:obevol}, the resonances have a vertical width $\Delta\alpha\approx 0.21''\cdot$yr$^{-1}$ for the $\phi_4$ resonance and $\Delta\alpha\approx 0.01''\cdot$yr$^{-1}$ for the $\phi_{19}$ resonance (computed at the right separatrix when it appears). Therefore, the time that would be needed for $\alpha$ to cross the $\phi_4$ resonance is $\Delta t\approx 2.5$~Gyrs, which corresponds to many oscillation periods of $\sigma_4$ (about $100$): this is the adiabatic regime. On the contrary, the time needed for $\alpha$ to cross the $\phi_{19}$ resonance is $\Delta t\approx 0.1$~Gyrs, which corresponds to less than one oscillation period of $\sigma_{19}$ (about $0.2$): this is the non-adiabatic regime. As a result, a resonance capture with $\phi_{19}$ is extremely unlikely, even if the orbital motion of Jupiter was restricted to its $19$th harmonic: Jupiter's spin axis enters the resonance and exits before $\sigma_{19}$ has time to oscillate. With suitable initial conditions, Jupiter roughly follows the resonance centre during the crossing, producing a bump in the obliquity evolution (see Fig.~\ref{fig:obevol} for $\lambda=0.227$), but nothing more can possibly happen. This kind of non-adiabatic resonance crossing is described by \cite{Ward-etal_1976}, \cite{Laskar-etal_2004b}, and \cite{Ward-Hamilton_2004} using Fresnel integrals.
   
\end{document}